# Tight lower bound of consecutive lengths for QC-LDPC codes with girth twelve

ZHANG GuoHua[1,2*] & WANG XinMei[1]

[1] *State Key Laboratory of Integrated Service Networks, Xidian University, Xi'an 710071, China;*

[2] *China Academy of Space Technology (Xi'an), Xi'an 710100, China*

*Corresponding author (email: zhangghcast@163.com)

**Abstract**: For an arbitrary (3,*L*) QC-LDPC code with a girth of twelve, a tight lower bound of the consecutive lengths is proposed. For an arbitrary length above the bound the resultant code necessarily has a girth of twelve, and for the length meeting the bound, the corresponding code inevitably has a girth smaller than twelve. The conclusion can play an important role in the proofs of the existence of large-girth QC-LDPC codes, the construction of large-girth QC-LDPC codes based on the Chinese remainder theorem, and the construction of LDPC codes with the guaranteed error correction capability.

**Key words**: low-density parity-check code, quasi-cyclic, girth, consecutive

Low-density parity-check (LDPC) codes have been the focus of intense research interest for channel coding community. In the construction of LDPC codes, the optimization objects most commonly used include the girth, the minimum distance, the minimum pseudo-codeword weight, and the minimum stopping set size and so on. By exploring the relations of these objects, researchers found that (when the parity-check matrix of an LDPC has a column weight of at least three) [1] the minimum distance grows exponentially with the girth, the minimum pseudo-codeword weight on the binary symmetric channel (BSC) for linear programming (LP) decoding increases exponentially with the girth, and the minimum stopping set size also increases exponentially with the girth. This conclusion suggests that many other objects can be significantly improved as the girth becomes larger. Besides, it is shown in [1] that the Gallager-A algorithm can correct all error patterns of weight $g/2-1$ in $g/2$ iterations for column-weight-three LDPC codes with girth $g \geq 10$. This implies that as far as the codes with the guaranteed error correction capability (GECC) are concerned, the error correction capability is largely depend on the value of girth.

Although many good algorithms [2-6] have been developed to construct large-girth LDPC codes, the code generated by these methods is generally fixed in the sense that if the length needs to be adjusted, these algorithms must restart from scratch. A technique distinct from these methods, is to construct large-girth LDPC codes with consecutive lengths [7,8]. The research of large-girth LDPC codes with consecutive lengths can play an important role [8] in some issues, such as the proofs of the existence of large-girth LDPC codes, the construction of large-girth LDPC codes based on the Chinese remainder theorem (CRT), and the construction of LDPC codes with GECC.

An LDPC code is defined as the null space of a sparse matrix. If the matrix is with uniform column weight of *R* and uniform row weight of *L*, then the resultant codes are termed as (*R,L*)-regular. If the matrix is composed of circulant permutation matrices (CPMs), then the corresponding codes are called quasi-cyclic (QC). In this paper, we denote a girth at least *g* by girth-$g^+$ and a girth equal to *g* by girth-*g*. Recently, a type of girth-$10^+$(3,*L*) QC-LDPC codes with consecutive lengths is proposed by [16] using the rings over finite polynomials. Such codes all have a girth at least ten, provided that the code lengths are larger than a certain lower bound; however, the method imposes some strict conditions on shift matrices and hence the lower bound can hardly cast light on designing general girth-$10^+$(3,*L*) QC-LDPC codes with consecutive lengths.

By investigating a general girth-$10^+$(3,*L*)QC-LDPC code defined by an arbitrary shift matrix, a tight lower bound of consecutive lengths for girth-$10^+$(3,*L*)QC-LDPC codes is discovered in [8].The tight lower bound can serve as a general guideline for designing girth-$10^+$(3,*L*)QC-LDPC codes with consecutive lengths, and naturally includes the lower bound of [7] as a special case.

It is well known that the maximal possible girth of a QC-LDPC code is twelve [9]; however, the consecutive lengths for girth-12 QC-LDPC codes are left as an open question in [8]. In this paper, we aim to answer this question: for a general girth-12 (3,*L*) QC-LDPC code, does a tight lower bound of consecutive lengths also exist? The main contribution of this paper is as follows. By analyzing the property of a general girth-12 (3,*L*) QC-LDPC code defined by an arbitrary shift matrix, a tight lower bound of consecutive lengths is discovered such that a (3,*L*)QC-LDPC code generated by the same shift matrix has a girth of twelve for an arbitrary length above the bound, and has a girth smaller than twelve for the length meeting the bound.

Since the length of a cycle is independent of the domain (binary or non-binary) of a parity-check matrix, the binary case is assumed below and the conclusion presented is also applicable to non-binary LDPC codes.

**1 Tight lower bound of consecutive lengths**

For a (3,*L*) QC-LDPC code with length *N=XL*, its parity-check matrix $\boldsymbol{H}_X$ can be expressed as [9]

$$\mathbf{H}_X = \begin{bmatrix} \mathbf{I}(0) & \mathbf{I}(0) & \cdots & \mathbf{I}(0) \\ \mathbf{I}(p_{1,0}) & \mathbf{I}(p_{1,1}) & \cdots & \mathbf{I}(p_{1,L-1}) \\ \mathbf{I}(p_{2,0}) & \mathbf{I}(p_{2,1}) & \cdots & \mathbf{I}(p_{2,L-1}) \end{bmatrix} \quad (1)$$

Where $\boldsymbol{I}(p)$ represents an $X \times X$ circulant permutation matrix with one at column-$(r+p)$ mod *X* for row-*r*, $0 \leq r \leq X-1$, and zero elsewhere.

The shift matrix $\boldsymbol{S}$ corresponding to $\boldsymbol{H}_X$ is denoted by

$$\mathbf{S} = \begin{bmatrix} 0 & 0 & \cdots & 0 \\ p_{1,0} & p_{1,1} & \cdots & p_{1,L-1} \\ p_{2,0} & p_{2,1} & \cdots & p_{2,L-1} \end{bmatrix} \quad (2)$$

where $p_{u,v} \in \{0,1,\ldots,X-1\}$ for $1 \leq u \leq 2, 1 \leq v \leq L-1$ and $p_{u,0}=0$. Without loss of generality, all elements in equ.(2) are assumed to be nonnegative, since negative integers can be turned into nonnegative ones by the operation of mod $X$.

$\mathbf{H}_X$ can be uniquely determined by $\mathbf{S}$ and $X$. Let $g(\mathbf{H}_X)$ be the girth of $\mathbf{H}_X$.

For $0 \leq j \leq L-1$, define $A=\max(p_{1,j})$, $B=\max(p_{2,j})$, $C=\max(p_{1,j}-p_{2,j})$ and $D=\max(p_{2,j}-p_{1,j})$, i.e., $A,B,C$ and $D$ are the maximal integers of the first line, of the second line, of the difference between the first and second lines, and of the difference between the second and first lines within $\mathbf{S}$, respectively. Let $T_1=2A+D$, $T_2=2B+C$, $T_3=A+C+2D$, $T_4=B+2C+D$, $T_5=A+B+D$, $T_6=A+B+C$ and $P'=\max\{T_1,T_2,T_3,T_4,T_5,T_6\}$.

***Lemma1***: Suppose $g(\mathbf{H}_Q)=12$ for a certain integer $Q$. Then $g(\mathbf{H}_P)=12$ for an arbitrary integer $P>P'$.

***Proof***: To prove lemma1, we need show that there exist no 4-, 6-,8- and 10-cycles within $\mathbf{H}_P$ (or equivalently within $\mathbf{S}$). Here are the hints. We first assume that there is a $t$-cycle ($t=4,6,8$ or $10$) within $\mathbf{H}_P$, then a formula holds in the modulus of $P$ due to equ.(4) of [9]. Then, by some basic algebraic operations, this formula is turned into a normal form such that its right-hand side (RHS) and left-hand side (LHS) are all nonnegative. Thus, if $P$ is larger than both the LHS and RHS, then the formula which holds in the modulus of $P$ also holds without the modulus. Therefore, the formula obviously holds in the modulus of an arbitrary integer, for example, $Q$, indicating a $t$-cycle within $\mathbf{H}_Q$, which contradicts $g(\mathbf{H}_Q)=12$.

First, we prove that $g(\mathbf{H}_P) \geq 10$ for an arbitrary integer $P>P'$, i.e., $\mathbf{H}_P$ has no 4-cycles, 6-cycles or 8-cycles. Since $T_1=2A+D \geq 2A$, $T_2=2B+C \geq 2B$ and $T_3=A+C+2D>C+C+2D$, we have $P' \geq 2\max\{A,B,C+D\}$. Consequently, by Lemma1 in [8], $g(\mathbf{H}_P) \geq 10$ for an arbitrary integer $P>P' \geq 2\max\{A,B,C+D\}$.

Next, we prove that $\mathbf{H}_P$ has no 10-cycles. If there were 10-cycles in $\mathbf{H}_P$, such cycles must appear as one of the three possible patterns shown in Fig.1, with other patterns all equivalent to one of the three ones (The detailed analysis is similar to the Appendix I in the online version of [8], and omitted here).

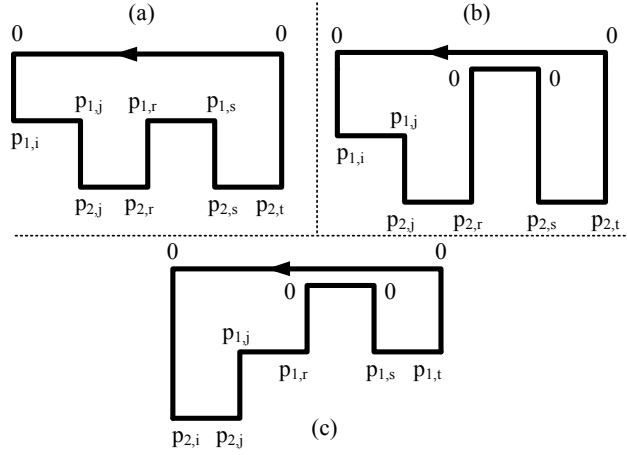

Fig.1 three possible 10-cycles patterns in $H_P$

Case(a): Assume that $H_P$ has a 10-cycle as Fig.1(a). Then there exist five positive integers $0 \leq i,j,r,s,t \leq L-1$ ($i \neq j; j \neq r, r \neq s; s \neq t; t \neq i$) such that

$$(0-p_{1,i})+(p_{1,j}-p_{2,j})+(p_{2,r}-p_{1,r})+(p_{1,s}-p_{2,s})+(p_{2,t}-0)=0 \pmod{P} \quad (3)$$

Equ.(3) can be expressed as

$$(p_{2,r}-p_{1,r})+p_{2,t}+p_{1,j}= p_{1,i}+p_{2,j}+(p_{2,s}-p_{1,s}) \pmod{P} \quad (4)$$

Case (a1): $(p_{2,r}-p_{1,r}) \geq 0$ and $(p_{2,s}-p_{1,s}) \geq 0$

Since the LHS and RHS of (4) meet $0 \leq (p_{2,r}-p_{1,r})+p_{2,t}+p_{1,j} \leq D+B+A$ and $0 \leq p_{1,i}+p_{2,j}+(p_{2,s}-p_{1,s}) \leq A+B+D$, respectively, we have $(p_{2,r}-p_{1,r})+p_{2,t}+p_{1,j}= p_{1,i}+p_{2,j}+(p_{2,s}-p_{1,s})$. A contradiction.

Case (a2): $(p_{2,r}-p_{1,r})<0$ and $(p_{2,s}-p_{1,s})<0$

In this case, (4) can be rewritten as

$$p_{2,t}+p_{1,j}+(p_{1,s}-p_{2,s})= p_{1,i}+p_{2,j}+(p_{1,r}-p_{2,r}) \pmod{P} \quad (5)$$

The LHS and RHS of (5) satisfy $0 \leq p_{2,t}+p_{1,j}+(p_{1,s}-p_{2,s}) \leq B+A+C$ and $0 \leq p_{1,i}+p_{2,j}+(p_{1,r}-p_{2,r}) \leq A+B+C$, respectively. Therefore, $p_{2,t}+p_{1,j}+(p_{1,s}-p_{2,s})= p_{1,i}+p_{2,j}+(p_{1,r}-p_{2,r})$. A contradiction.

Case (a3): $(p_{2,r}-p_{1,r}) \geq 0$ and $(p_{2,s}-p_{1,s})<0$

This case includes two subcases.

Subcase (a3-1): $(p_{1,j}-p_{2,j}) \geq 0$

In this case, (4) can be expressed as

$$(p_{2,r}-p_{1,r})+p_{2,t}+(p_{1,j}-p_{2,j})+(p_{1,s}-p_{2,s})= p_{1,i} \pmod{P} \quad (6)$$

The LHS and RHS of (6) satisfy $0 \leq (p_{2,r}-p_{1,r})+p_{2,t}+(p_{1,j}-p_{2,j})+(p_{1,s}-p_{2,s}) \leq D+B+C+C$ and $0 \leq p_{1,i} \leq A$, respectively. Thus, $(p_{2,r}-p_{1,r})+p_{2,t}+(p_{1,j}-p_{2,j})+(p_{1,s}-p_{2,s})= p_{1,i}$. A contradiction.

Subcase (a3-2): $(p_{1,j}-p_{2,j})<0$

In this case, (4) can be rewritten as

$$(p_{2,r}-p_{1,r})+p_{2,t}+(p_{1,s}-p_{2,s})= p_{1,i}+(p_{2,j}-p_{1,j}) \pmod{P} \quad (7)$$

The LHS and RHS of (7) satisfy $0 \leq (p_{2,r}-p_{1,r})+p_{2,t}+(p_{1,s}-p_{2,s}) \leq D+B+C$ and $0 \leq p_{1,i}+(p_{2,j}-p_{1,j}) \leq A+D$, respectively. Consequently, $(p_{2,r}-p_{1,r})+p_{2,t}+(p_{1,s}-p_{2,s})= p_{1,i}+(p_{2,j}-p_{1,j})$. A contradiction.

Case (a4): $(p_{2,r}-p_{1,r})<0$ and $(p_{2,s}-p_{1,s}) \geq 0$

This case also includes two sub-cases.

Subcase (a4-1): $(p_{1,j}-p_{2,j})\geq 0$

In this case,(4) can be expressed as

$$p_{2,t}+(p_{1,j}-p_{2,j})= p_{1,i}+(p_{1,r}-p_{2,r})+(p_{2,s}-p_{1,s}) \pmod{P} \quad (8)$$

The LHS and RHS of (8) meet $0\leq p_{2,t}+(p_{1,j}-p_{2,j})\leq B+C$ and $0\leq p_{1,i}+(p_{1,r}-p_{2,r})+(p_{2,s}-p_{1,s})\leq A+C+D$, respectively. Therefore, $p_{2,t}+(p_{1,j}-p_{2,j})=p_{1,i}+(p_{1,r}-p_{2,r})+(p_{2,s}-p_{1,s})$. A contradiction.

Subcase (a4-2): $(p_{1,j}-p_{2,j})<0$

In this case, (4) can be expressed as

$$p_{2,t}= p_{1,i}+(p_{1,r}-p_{2,r})+(p_{2,s}-p_{1,s})+(p_{2,j}-p_{1,j}) \pmod{P} \quad (9)$$

The LHS and RHS of (9) satisfy $0\leq p_{2,t}\leq B$ and $0\leq p_{1,i}+(p_{1,r}-p_{2,r})+(p_{2,s}-p_{1,s})+(p_{2,j}-p_{1,j})\leq A+C+2D$, respectively. Thus, $p_{2,t}= p_{1,i}+(p_{1,r}-p_{2,r})+(p_{2,s}-p_{1,s})+(p_{2,j}-p_{1,j})$. A contradiction.

Case (b): Assume that $H_P$ has a 10-cycle as Fig.1(b). Then there exist five positive integers $0\leq i,j,r,s,t\leq L-1$ ($i\neq j;j\neq r,r\neq s;s\neq t;t\neq i$) such that

$$(0-p_{1,i})+(p_{1,j}-p_{2,j})+(p_{2,r}-0)+(0-p_{2,s})+(p_{2,t}-0)=0 \pmod{P} \quad (10)$$

Case (b1): $(p_{2,j}-p_{1,j})\geq 0$.

In this case,(10) can be rewritten as

$$p_{2,r}+p_{2,t} = p_{2,s}+p_{1,i}+(p_{2,j}-p_{1,j}) \pmod{P} \quad (11)$$

Since the LHS and RHS of (11) meet $0\leq p_{2,r}+p_{2,t}\leq 2B$ and $0\leq p_{2,s}+p_{1,i}+(p_{2,j}-p_{1,j})\leq B+A+D$, respectively, we have $p_{2,r}+p_{2,t} = p_{2,s}+p_{1,i}+(p_{2,j}-p_{1,j})$. A contradiction.

Case (b2): $(p_{2,j}-p_{1,j})<0$

In this case, (10) can be expressed as

$$p_{2,r}+p_{2,t}+(p_{1,j}-p_{2,j})= p_{2,s}+p_{1,i} \pmod{P} \quad (12)$$

Since the LHS and RHS of (12) meet $0\leq p_{2,r}+p_{2,t}+(p_{1,j}-p_{2,j})\leq 2B+C$ and $0\leq p_{2,s}+p_{1,i}\leq B+A$, respectively, we have $p_{2,r}+p_{2,t}+(p_{1,j}-p_{2,j})= p_{2,s}+p_{1,i}$. A contradiction.

Case (c): Assume that $H_P$ has a 10-cycle as Fig.1(c). Then there exist five positive integers $0\leq i,j,r,s,t\leq L-1$ ($i\neq j;j\neq r,r\neq s;s\neq t;t\neq i$) such that

$$(0-p_{2,i})+(p_{2,j}-p_{1,j})+(p_{1,r}-0)+(0-p_{1,s})+(p_{1,t}-0)=0 \pmod{P} \quad (13)$$

Case(c1): $(p_{2,j}-p_{1,j})\geq 0$

In this case, (13) can be expressed as

$$(p_{2,j}-p_{1,j})+p_{1,r}+p_{1,t}=p_{1,s}+p_{2,i} \pmod{P} \quad (14)$$

Since the LHS and RHS of (14) satisfy $0\leq (p_{2,j}-p_{1,j})+p_{1,r}+p_{1,t}\leq D+A+A$ and $0\leq p_{1,s}+p_{2,i}\leq A+B$, respectively, we have $(p_{2,j}-p_{1,j})+p_{1,r}+p_{1,t}=p_{1,s}+p_{2,i}$. A contradiction.

Case(c2): $(p_{2,j}-p_{1,j})<0$

In this case, (13) can be rewritten as

$$p_{1,r}+p_{1,t}=p_{1,s}+p_{2,i}+(p_{1,j}-p_{2,j}) \pmod{P} \quad (15)$$

Since the LHS and RHS of (15) satisfy $0 \leq p_{1,r}+p_{1,t} \leq A+A$ and $0 \leq p_{1,s}+p_{2,i}+(p_{1,j}-p_{2,j}) \leq A+B+C$, respectively, we have $p_{1,r}+p_{1,t}=p_{1,s}+p_{2,i}+(p_{1,j}-p_{2,j})$. A contradiction.

<div align="right">Q.E.D.</div>

**Lemma2:** $g(H_{P'})<12$.

**Proof**: Denote $\arg\{\max(a_j)|0 \leq j \leq L-1\}$ as the subscript of the maximal value of a sequence of $L$ integers: $a_0,a_1,\ldots,a_{L-1}$. Let $k=\arg\{\max(p_{2,j}-p_{1,j})|0 \leq j \leq L-1\}$, $r=\arg\{\max(p_{1,j})|0 \leq j \leq L-1\}$, $s=\arg\{\max(p_{2,j})|0 \leq j \leq L-1\}$ and $t=\arg\{\max(p_{1,j}-p_{2,j})|0 \leq j \leq L-1\}$. We consider the six possible cases separately.

Case (1) $P'=T_1$:

*Proof*: Obviously, $r \neq 0$. Assume $k=r$. Then $A<B$, and hence $T_1=2A+D<A+B+D=T_5$. A contradiction. Thus, we have $k \neq r$.

Case (1.1) $k=0$: $P'=2A+D=2A+0$. Hence, $H_{P'}$ has an 8-cycle described by

$$(p_{0,0}-p_{1,0})+(p_{1,r}-p_{0,r})+(p_{0,0}-p_{1,0})+(p_{1,r}-p_{0,r})=2A=0 \pmod{P'} \quad (16)$$

Case (1.2) $k \neq 0$: Since $0 \neq r$, $0 \neq k$ and $k \neq r$, $H_{P'}$ has a 10-cycle described by

$$(p_{0,0}-p_{1,0})+(p_{1,r}-p_{0,r})+(p_{0,0}-p_{2,0})+(p_{2,k}-p_{1,k})+(p_{1,r}-p_{0,r})=0+A+0+D+A=0 \pmod{P'} \quad (17)$$

In summary, when $P'=T_1$, $g(H_{P'})<12$.

Case (2) $P'=T_2$:

*Proof*: Similarly to Case (1), we can prove that $g(H_{P'})<12$ for Case (2).

Case (3) $P'=T_3$: $T_3=A+C+2D$

*Proof*: Obviously, $r \neq 0$ and $k \neq t$. First, assume $t=0$. Then $C=0$, and hence $T_3=A+2D<A+B+D=T_5$. A contradiction. Therefore, $t \neq 0$. Next, assume $k=0$. Then $D=0$ and hence $T_3=A+C<A+B+C=T_6$. A contradiction. Thus, $k \neq 0$. Again, assume $k=r$. Then $T_3=C+(A+2D)<C+2(A+D) \leq C+2B=T_2$. A contradiction. Therefore, $k \neq r$.

Since $k \neq 0, k \neq t, k \neq r$ and $r \neq 0$, $H_{P'}$ has a 10-cycle described by

$$(p_{0,0}-p_{2,0})+(p_{2,k}-p_{1,k})+(p_{1,t}-p_{2,t})+(p_{2,k}-p_{1,k})+(p_{1,r}-p_{0,r})=0+D+C+D+A=0 \pmod{P'} \quad (18)$$

In summary, when $P'=T_3$, $g(H_{P'})<12$.

Case (4) $P'=T_4$:

*Proof*: Similarly to Case (3), we can prove that $g(H_{P'})<12$ for Case (4).

Case (5) $P'=T_5$:

Obviously, $r \neq 0$ and $s \neq 0$. Assume $k=0$. Then $D=0$ and hence $C>0$. Therefore, $T_5=A+B+D<A+B+C=T_6$. A contradiction. Thus, $k \neq 0$.

Case (5.1) $r=k$: Since $T_5=A+B+D=B+(D+A)=B+B$, $H_{P'}$ has an 8-cycle described by

$$(p_{0,0}-p_{2,0})+(p_{2,s}-p_{0,s})+(p_{0,0}-p_{2,0})+(p_{2,s}-p_{0,s})=2B=0 \pmod{P'} \quad (19)$$

Case (5.2) $r \neq k$: Since $s \neq 0, k \neq 0, k \neq r$ and $r \neq 0$, $H_{P'}$ has a 10-cycle described by

$$(p_{0,0}-p_{2,0})+(p_{2,s}-p_{0,s})+(p_{0,0}-p_{2,0})+(p_{2,k}-p_{1,k})+(p_{1,r}-p_{0,r})=0+B+0+D+A=0 \pmod{P'} \quad (20)$$

In summary, when $P'=T_5$, $g(H_{P'})<12$.

Case (6) $P'=T_6$:

*Proof*: Similarly to Case (5), we can prove that $g(H_{P'})<12$ for Case (6).

To sum up, $g(\mathbf{H}_{P'})<12$ always holds for whatever cases.

Q.E.D.

From Lemmas 1 and 2, the main result of this paper can be described below:

**Theorem 1:** Suppose $g(\mathbf{H}_Q)=12$ for a certain integer $Q$. Then $N=LP'$ is a tight lower bound such that $g(\mathbf{H}_P)=12$ for all lengths above the bound (i.e., $P>P'$), and $g(\mathbf{H}_P)<12$ for the length meeting the bound (i.e., $P=P'$), where $P'=\max\{T_1,T_2,T_3,T_4,T_5,T_6\}$, $T_1=2A+D$, $T_2=2B+C$, $T_3=A+C+2D$, $T_4=B+2C+D$, $T_5=A+B+D$ and $T_6=A+B+C$. $A,B,C$ and $D$ are the maximal integers of the first line, of the second line, of the difference between the first and second lines, and of the difference between the second and first lines within $\mathbf{S}$, respectively.

## 2 Some applications of the tight lower bound

The existence issue of girth-12 (3,$L$) QC-LDPC codes is a very hard topic. Generally speaking, the longer the code length, the more the chance of girth-12 conditions being satisfied for a (3,$L$) QC-LDPC code. On the other hand, the shorter the code length, the less the chance. Therefore, the literatures available all tend to illustrate the efficiency of their methods by record the shortest girth-12 LDPC code ever found [5]. Since such methods all aim at a specific block length when searching or constructing a code, they only provide some scattered results (i.e., find a girth-12 code for a single length or several codes for a few specific lengths)[5,6]. Distinct from these methods, by Theorem 1 we can readily obtain some conclusions on the existence issue for consecutive lengths. Based on Theorem 1, if there exists a 3×$L$ shift matrix $\mathbf{S}$ such that $g(\mathbf{H}_Q)=12$ for a certain integer $Q$, then all girth-12 (3,$L$) QC-LDPC codes with lengths $N=PL(P\geq\max\{T_1,T_2,T_3,T_4,T_5,T_6\}+1)$ do exist and hence their existence issue are completely settled.

**Example 1:** We developed a search method based on simulated annealing [10]. By this method, we found a 3×6 shift matrix $\mathbf{S}$ as shown in equ.(21).

$$\mathbf{S} = \begin{bmatrix} 0 & 0 & 0 & 0 & 0 & 0 \\ 0 & 3 & 14 & 18 & 24 & 26 \\ 0 & 19 & 62 & 107 & 170 & 224 \end{bmatrix} \quad (21)$$

It is readily verified that $g(\mathbf{H}_Q)=12$ for $Q=393$. According to Theorem 1, $g(\mathbf{H}_P)=12$ for all the CPM sizes $P\geq 2\times 244+1=449$. This is an encouraging result considering that the up-to-date shortest girth-12 (3,6) QC-LDPC code, to the best of our knowledge, is reported with length 2874 by O'Sullivan [5] and that equ.(21) combined with Theorem 1 can explicitly give an infinite set of girth-12 (3,6) QC-LDPC codes with consecutive lengths (with 2694 as the starting length and an incremental step of six). Clearly, the infinite set include 30 member codes with consecutive lengths (with a step of six), which are even shorter than length (2874) of the shortest girth-12 (3,6) QC-LDPC codes presented by O'Sullivan.

The above example suggests that the tight lower bound newly proposed can find important application in the existence issue of girth-12 QC-LDPC codes. Just like the application scope and significance of the results in [8], the results obtained in this paper should be also useful for the construction of large-girth QC-LDPC codes based on the CRT, the construction of LDPC codes with GECC and some related areas of cryptography. For these applications of the tight lower bound, the interested readers can see section 2 of [8].

3 Conclusions

In the optimization of LDPC codes, girth is one of the most significant factors. It is well-known that the maximal girth of a QC-LDPC code is twelve [9]. For general girth-12($3,L$)QC-LDPC codes, a new important property is discovered and proved in this paper, which states that given a girth-12($3,L$)QC-LDPC code, its shift matrix and an arbitrary CPM size $P$ ($P$ above a threshold determined by the shift matrix) necessarily produce a QC-LDPC code with a girth of twelve. The threshold is also proved to be optimal in the sense that when $P$ equals the threshold the corresponding QC-LDPC code inevitably has a girth less than twelve. Therefore, the important open problem unsolved in [8], i.e, the consecutive lengths for QC-LDPC codes with the maximal girth of twelve, is completely settled in this paper.

*This work was supported by the National Basic Research Program of China (2010CB328300), the National Natural Science Foundation of China (61001130,61001131) and "111" Project (B08038).*

1  Chilappagari S K, Nguyen D V, Vasic B,et al.Error correction capability of column-weight-three LDPC codes under the Gallager A Algorithm-part II. IEEE Trans Inf Theory,2010,56(6):2626—2639
2  Fujisawa M, Sakata S. A construction of high rate quasi-cyclic regular LDPC codes from cyclic difference families with girth 8.IEICE Trans Fund El Comm Comp Sci, 2007,E90-A:1055—1061
3  Tao X F, Kim J M, Liu W Z, et al. Improved construction of low-density parity-check codes based on lattices.ISITC2007, Jeonju, Korea,2007,208—212
4  Zhang F, Mao X H, Zhou W Y, et al.Girth-10 LDPC codes based on 3-D cyclic lattices.IEEE Trans Veh Techn, 2008,57(2):1049—1060
5  O'Sullivan M E.Algebraic construction of sparse matrices with large girth.IEEE Trans Inf Theory,2006,52(2):718—727
6  Milenkovic O,Kashyap N,Leyba D. Shortened array codes of large girth.IEEE Trans Inf Theory, 2006, 52(8):3707—3722
7  Liu L, Zhou W Y. Design of QC-LDPC code with continuously variable length (in Chinese). J El Inf Tech, 2009,31(10):2523—2526.
8  Zhang G H, Wang J H, Li X Y, et al. Tight lower bound of consecutive lengths for QC-LDPC codes with girth at least ten. Chinese Sci Bull,2011,56(12):1272—1277.

9 Fossorier M P C.Quasi-cyclic low-density parity-check codes from circulant permutation matrices.IEEE Trans Inf Theory,2004,50(8):1788—1793

10 Zhang G H, Wang X M, Zhou Q. Construction of girth-12(3,5)- and (3,6)-regular QC-LDPC codes based on simulated annealing algorithm (in Chinese). Space Electronic Technology, 2010,7(3): 69—73

11 Zhou L, Li S Q. New direction for joint design of stream cipher and error-correcting codes—advances of research on fast correlation attack decoding algorithm. Journal of University of Electronic Science and Technology of China, 2009, 38(5):555—561.